\documentclass[acmsmall, screen]{acmart}
\usepackage{subcaption}
\usepackage[most]{tcolorbox}
\usepackage{lipsum}
\usepackage{booktabs}
\usepackage{array}
\usepackage{hyperref}
\usepackage{enumitem}
\usepackage{soul} 

\newif\ifshowchanges
\showchangestrue
\showchangesfalse 

\ifshowchanges
    \usepackage{changes} 
\else 
    
    \newcommand{\deleted}[1]{}

\fi
\usepackage{tikz}
 
\usepackage{listings}
\AtBeginDocument{%
  \providecommand\BibTeX{{%
    \normalfont B\kern-0.5em{\scshape i\kern-0.25em b}\kern-0.8em\TeX}}}

\usepackage{graphicx}
\usepackage{longtable}
\usepackage{multirow}

\setcopyright{acmcopyright}
\copyrightyear{2018}
\acmYear{2018}
\acmDOI{XXXXXXX.XXXXXXX}

\acmConference[Conference acronym 'XX]{Make sure to enter the correct
  conference title from your rights confirmation emai}{June 03--05,
  2018}{Woodstock, NY}
\acmISBN{978-1-4503-XXXX-X/18/06}

\begin{document}
\title[Adapting to LLMs]{Adapting to LLMs: How Insiders and Outsiders Reshape Scientific Knowledge Production} 


\author{Huimin Xu}
\affiliation{%
  \institution{School of Information, University of Texas at Austin}
  \city{Austin}
  \state{Texas}
  \country{USA}
}

\author{Houjiang Liu}
\affiliation{%
  \institution{School of Information, University of Texas at Austin}
  \city{Austin}
  \state{Texas}
  \country{USA}
}

\author{Yan Leng}
\affiliation{%
  \institution{McCombs School of Business, University of Texas at Austin}
  \city{Austin}
  \state{Texas}
  \country{USA}
}

\author{Ying Ding}
\affiliation{%
  \institution{School of Information, University of Texas at Austin}
  \city{Austin}
  \state{Texas}
  \country{USA}
}




\begin{abstract}

CSCW has long examined how emerging technologies reshape the ways researchers collaborate and produce knowledge, with scientific knowledge production as a central area of focus. As AI becomes increasingly integrated into scientific research, understanding how researchers adapt to it reveals timely opportunities for CSCW research --- particularly in supporting new forms of collaboration, knowledge practices, and infrastructure in AI-driven science.

This study quantifies LLM impacts on scientific knowledge production based on an evaluation workflow that combines an insider-outsider perspective with a knowledge production framework. 
Drawing on the OpenAlex dataset, we identify 7,106 researchers as insiders or outsiders based on their involvement in LLM development. We then analyze their publications using few-shot prompting to reveal shifts in knowledge production, demonstrating how LLMs reshape their research directions. To ensure robustness, we further validate these findings with bibliometric measurements.
Our findings reveal how LLMs catalyze both innovation and reorganization in scientific communities, offering insights into the broader transformation of knowledge production in the age of generative AI and sheds light on new research opportunities in CSCW.

\end{abstract}



\maketitle

\section{Introduction} \label{sec:Introduction}

\begin{quote}
    ``It is not the strongest of the species that survives, nor the most intelligent that survives. It is the one that is most adaptable to change.'' - Charles Darwin
\end{quote}

Darwin's quote, rooted in the natural world, resonates strongly with the challenges facing both companies and universities today. In rapidly changing environments, adaptability, the ability to evolve and innovate, serves as a key driver of technological and scientific advancement. One of the most pressing challenge organizations face is to balance the imperative to produce with the need to innovate in response to change \citep{Uhl-Bien2018-vk}. This tension is often framed as a trade-off between exploitation (maximizing existing strengths) and exploration (pursuing novel opportunities). As \citet{Uhl-Bien2018-vk} argue, the most resilient organizations are those that can dynamically integrate both modes of operation.

The rise of generative AI and large language models (LLMs) has accelerated the pace for scientists to adapt their research portfolios in response to emerging opportunities and challenges. Scientists who previously relied on traditional or disciplinary theories and methods have begun incorporating LLMs and large-scale data analytics into their established work. This shift is not merely technical; it reflects a deeper cognitive adaptability, a willingness to learn new tools, critically reflect on possible issues, rethink long-standing research paradigms, and envision the future scientific knowledge production. 

LLMs are rapidly reshaping what is possible across disciplines, from accelerating discovery in the natural sciences \citep{Luo2025-ji} to transforming methodologies in the social sciences and humanities \citep{Bail2024-pf}. In natural science, LLMs have outperformed experts in predicting neuroscience experimental results \citep{Luo2025-ji}. In social science, \citet{Bail2024-pf} claimed that ``generative AI has the potential to improve survey research, online experiments, automated content analyses, agent-based models, and other techniques commonly used to study human behavior.'' LLMs not only bring tremendous potential and applications for scientists, they have also raised new reflections about ethics, fairness, hallucination, reproducibility, and plagiarism \citep{Bail2024-pf, Stokel-Walker2023-dm, otis2024global}.

Aligning the adaptive phenomenon of scientists, we propose the central research question: \textit{How are scientists adapting their research practices in response to LLMs, and what does this reveal about the broader transformation of scientific knowledge production in the age of generative AI?} Although many researchers now use LLMs to support or automate parts of their scientific workflows, or investigate LLMs as research subjects themselves, large-scale empirical evaluations of AI’s broader impact on science remain limited \citep{Ackerman2013-vc, Correia2019-yi, Jirotka2013-gt, Neang2021-qi}. Most current assessments rely on individual experiences or narrow, domain-specific experiments, rather than systematic, cross-disciplinary studies. This urges a need for a comprehensive framework to understand how AI is transforming scientific practice.

Toward this end, we develop an evaluation workflow to explain how LLMs are reshaping scientific knowledge production. Our workflow builds on the insider–outsider classification of radical innovation \citep{Hill2003-jc, Van-De-Poel2000-fa}, integrated with a comparative framework of knowledge production modes \citep{Scott2012-ub, Hessels2008-ew}. In the context of LLM-driven scientific knowledge production, we first distinguish between ``insiders'', such as computer scientists and AI researchers, who are directly involved in building, fine-tuning, and advancing LLM technologies, and ``outsiders'', who apply LLMs within specific domain contexts, such as medicine, biology, or psychology, often without direct involvement in the technical development. We then examine research publications based on the knowledge production framework, which identifies key differences between \textit{two modes of knowledge production across five dimensions}: application focus, disciplinary orientation, collaboration patterns, social accountability, and evaluation criteria. 

To quantify the impact of LLMs based on the evaluation workflow above, we conduct a computational analysis that combines few-shot classification powered by LLM with bibliometric validation. Specifically, using the OpenAlex dataset \citep{Priem2022-xb}, we identify 7,106 researchers, classified as insiders or outsiders, who have published LLM-related work, along with their prior publications. We then classify research abstracts from these scholars along \textit{five knowledge dimensions} using LLM few-shot prompting, enabling us to assess researchers' knowledge production shifts before and after LLM adoption. To ensure robustness, we validate the LLM-generated classifications through bibliometric techniques, providing quantitative support for the observed patterns and trends.

Our findings reveal that the adoption of LLMs is reshaping scientific knowledge production in distinct ways. Outsiders, researchers from non-AI domains, are increasingly leveraging LLMs to pursue application-focused research, engage in transdisciplinary work, address social accountability, and experiment with new evaluation practices. In response, insiders are restructuring their collaboration networks, forming partnerships across a broader range of institutional contexts. These shifts highlight timely opportunities for CSCW research: to support outsider-led integration of LLMs through the design of domain-specific tools, and to investigate emerging mediator roles that facilitate transdisciplinary collaboration, data ownership, and shared infrastructure in AI-driven science.

\section{Related Work} \label{sec:Background}

\subsection{The Role of AI in Scientific Knowledge Production} \label{subsec:AI roles}

CSCW has long examined how technologies shape the ways researchers collaborate and produce knowledge, with scientific knowledge production being a key area of focus \citep{Ackerman2013-vc, Correia2019-yi, Jirotka2013-gt, Neang2021-qi}. With the increasing integration of AI technologies into research workflows, new opportunities and challenges emerge. In particular, we identify two prominent lines of inquiry concerning the role of AI in scientific knowledge production and collaboration.

\subsubsection{AI as a core infrastructure that mediates and transforms scientific collaborative work} Recent advances in LLMs have accelerated the development of new tools that assist researchers in tasks such as literature synthesis, idea generation, data analysis, and academic writing. These tools include open-source ones like ScholarQA \citep{Asai2024-sz}, which helps researchers synthesize literature and identify research gaps and CollabCoder \citep{Gao2024-ku}, which supports inductive collaborative qualitative analysis. An increasing number of new tools are being introduced. Alongside these efforts, commercial tools like ChatGPT, Perplexity, Elicit, Consensus, and others also achieved widespread adoption, further embedding LLMs into the everyday practices of research \citep{Liang2024-vz, Naddaf2025-yz}. 

Beyond general-purpose tools, domain-specific LLMs are being fine-tuned to exceed human-level reasoning within specialized fields. For example, \citet{Luo2025-ji} finetune LLMs on neuroscience literature, demonstrating that they could outperform expert researchers in predicting experimental outcomes. \citet{Sun2024-bn} developed Mephisto, a multi-agent framework for interpreting multi-band galaxy observations, automating complex tasks in astrophysics. Similarly, \citet{Ghafarollahi2024-in} develops SciAgents, which employs agentic workflows to solve mathematical problems. These examples underscore how AI is not merely supporting scientific workflows but increasingly shaping how scientific knowledge is created.

\subsubsection{AI itself as an object of research investigation}
The second line of inquiry shifts the focus toward AI itself, treating it as an object of research investigation and contextualizing it through the lens of researchers' disciplinary expertise. Here, researchers from diverse disciplinary backgrounds come together to interpret, adapt, and critique emerging AI technologies. Again, using LLMs as an example, \citet{Pang2025-eo} reviewed LLM-related studies published at CHI from 2020 to 2024, revealing a growing body of work focused on interrogating the mechanisms and limitations of LLMs, including issues of training data provenance, response behavior, and persistent risks such as hallucination and misinformation. 

Beyond the HCI community, scholars in other disciplines have begun to interrogate LLMs through the lens of their own domain expertise. In psychology, for example, \citet{Hagendorff2023-gu} introduced the notion of machine psychology and proposed to analyze how LLMs replicate or diverge from human cognitive heuristics, biases, and social reasoning abilities. Building on this proposal, \citet{Almeida2024-ll} applied classic psychological experiments to LLMs, probing their susceptibility to phenomena like hindsight bias, rule violations, deception, and other human cognitive tasks. These studies reflect an important epistemic shift: AI is not only a tool for scientific inquiry but also a research subject, prompting domain-specific inquiries into how its human-like capabilities might align with or challenge established theories and disciplinary assumptions.

\vspace{0.2cm}
\noindent Together, these two lines of inquiry highlight how scientific knowledge production is both mediated by AI technologies and increasingly oriented around them. Research teams do more than adopt AI as infrastructure, they actively negotiate its meaning, reconcile disciplinary values, and co-construct new epistemic norms \citep{Wang2023-hl}. Thus, we argue that this dual role of AI, as both collaborator in research and an object of inquiry, reconfigures the dynamics of interdisciplinary collaboration and calls for new theoretical frameworks to understand and evaluate its impact on scientific knowledge production.

\subsection{Evaluating AI Impact in Scientific Knowledge Production} \label{subsec:Evaluation}

Two main approaches are commonly observed in evaluating AI's impact on scientific knowledge production: \textit{micro-level evaluations}, and \textit{macro-level evaluations}. Notably, these two evaluation approaches are reflected in both lines of inquiry described in Section \ref{subsec:AI roles}, reflecting the localized and systemic dimensions of AI integration in science.


%



\subsubsection{Micro-level evaluations} Researchers who develop and deploy LLM-powered tools, whether for general research activities \citep{Asai2024-sz, Gao2024-jw}, such as literature synthesis \citep{Lee2024-bb, Chan2018-of, Kang2022-we, Kang2023-yh} and research ideation \citep{KapaniaShivani2025-wu, Liu2024-jo}, or domain-specific tasks \citep{Ghafarollahi2024-in, Luo2025-ji, Sun2024-bn}, typically evaluate these systems based on metrics such as task efficiency, usability, or user satisfaction. While these criteria provide valuable insights into immediate tool performance, they offer limited understanding of how AI may reshape deeper epistemic practices, research workflows, or disciplinary knowledge structures. Most current evaluations of LLM's impact on scientistic practice remain grounded in individual-level experiences and self-reported perceptions \citep{Binz2025-bz}. 

Similarly, many domain experts who explore LLMs through the lens of their disciplinary expertise also adopt a micro-level orientation, focusing on concerns such as hallucination, bias, or alignment with human reasoning in specific contexts \citep{Almeida2024-ll, Hagendorff2023-gu, Pang2025-eo}. These studies also treat LLMs as standalone entities, paying less attention to the complex scientific workflows and interdisciplinary collaborations in which these tools are embedded. In this case, the longer-term implications for collective knowledge production influenced by AI are less frequently examined. 




\subsubsection{Macro-level evaluations} A growing body of large-scale empirical research is beginning to systematically evaluate the broad benefits and limitations of AI in scientific knowledge production. These studies examine multiple dimensions of impact. One key focus is on \textit{productivity}. \citet{Liang2024-vz} analyzed 950,965 papers published between 2020 and 2024 to measure the prevalence of LLM-modified content. Their findings revealed a significant increase in LLM usage, particularly in Computer Science, where up to 17.5\% of papers showed signs of AI-generated modifications. The study also observed that researchers who frequently publish preprints tend to be early adopters of LLM tools. 

Beyond productivity, AI adoption also influences the \textit{visibility} and \textit{diffusion} of scientific work. In a large-scale bibliometric study of 87.6 million papers, \citet{Gao2024-jw} found that papers incorporating AI technologies were more likely to be cited, both within and across disciplinary boundaries. These findings suggest that LLMs are beginning to shape not only how research is conducted, but also how it circulates and gains recognition in the broader scientific ecosystem. \textit{Fairness} in AI adoption is another emerging area of concern. A meta-analysis of 18 studies involving over 140,000 participants worldwide found that women are generally underrepresented in the use of generative AI tools such as OpenAI's ChatGPT and Anthropic's Claude \citep{otis2024global}. Within the scientific workforce, \citet{Gao2024-jw} further identified disparities across disciplines, noting that fields with higher proportions of underrepresented groups, such as women and Black scholars, tend to benefit less from the citation advantages associated with AI adoption. These patterns highlight the need for more inclusive and equitable approaches as AI becomes increasingly embedded in research practices.

\vspace{0.2cm}
\noindent By examining both micro- and macro-level evaluations of AI’s impact on scientific knowledge production, we find that existing research is fragmented across disciplines and pays limited attention to the socially situated and collaborative practices emphasized in CSCW. While much of the current work focuses on tool performance or large-scale systemic outcomes, it often neglects the ways in which scientific collaboration, disciplinary positionality, and social dynamics among academic institutions influence who contributes to knowledge production, how new practices are negotiated, and the conditions under which epistemic change occurs \citep{Correia2019-yi, Jirotka2013-gt}.

To address this gap, we draw on the insider–outsider framework to better understand how researchers’ positions within or outside dominant disciplinary and institutional structures influence their capacity to shape AI-driven transformations. In the next section, we introduce related work that informs this perspective.



\subsection{Insider vs. Outsider Adaptive Changes } \label{subsec:Insider-outsider}

Adapting to changes is essential, but insiders and outsiders often exhibit different patterns in how they respond to such changes. Across domains, research has shown that those outsiders who are not deeply embedded within the dominating groups may bring novel ideas, different connections and even challenge the status quo \citep{BarsouxUnknown-eh}.

In \textit{The Innovator’s Dilemma}, \citet{Christensen2015-ve} claimed that incumbent companies, even they are very successful, usually fail to adopt disruptive innovations since it will undermine their existing business models. These technological innovations are more likely to be driven and embraced by companies who are outsiders since they are constrained by existing customer demands and sunk costs. 

This idea can also be extended in scientific domain. \citet{Kuhn2012-zv} states that revolutionary science and paradigm shifts tend to emerge from outside of the dominant groups, whereas insiders often continue the normal science. Bourdieu’s theory \citep{Bourdieu1975-ll} regarding scientific field also suggests that individuals with less capitals in the dominant groups (outsiders) are more likely to challenge the existing norms and established authority. 

Empirical studies support this idea. In an experiment by \citet{Franke2014-el}, groups of carpenters, roofers, and inline skaters were asked to improve safety equipment for domains both inside and outside their own expertise. Each group generated more novel solutions for the domains outside their own, suggesting that distance from established routines can foster more creative thinking. Taken together, these findings suggest that insiders tend to favor incremental innovations that align with existing knowledge and norms, while outsiders are more likely to pursue disruptive innovations as a strategy to enter or reshape a field.



\subsubsection{Benefits of outsider-within for scientific knowledge production}
A wide range of literature offers explanations for why outsiders are often better positioned to drive radical innovation. One widely accepted explanation is that \textit{outsiders are typically free from the organizational inertia, risk aversion, and path dependencies} that constrain incumbents \citep{Hill2003-jc}. Unlike insiders who may be restricted by existing customer markets, resource allocations, or internal routines within organizations, outsiders can explore new ideas in niche markets without needing to justify them within an established structure. 

A second explanation emphasizes outsiders' freedom from disciplinary conventions. \citet{Van-De-Poel2000-fa} suggests that outsiders may benefit because they are not constrained by the rules or conventions of an established technological field. Their lack of embeddedness enables more flexible thinking and risk-taking. Both explanations highlight outsiders' ability to innovate more freely because they are not embedded in traditional systems or conventions. \citet{Hill2003-jc}'s explanation focuses more on organizational inertia and internal constraints, while \citet{Van-De-Poel2000-fa} adds that outsiders are freer from rules and conventions of the established field. 

A third explanation is, outsiders often draw on diverse social capitals \citep{Smith2012-cp}, allowing them to integrate ideas across domains and build unconventional collaborations, an especially valuable asset when adopting or adapting new technologies. By illustrating a single innovation that transformed the technology of Formula 1 motor racing, \citet{Smith2012-cp} showed that how outsiders could outperform the existing insiders by taking advantage their informal networks and weak ties. The social capital resources of outsiders place them at an advantage in adapting to new technologies. 

\subsubsection{Challenges of outsider-within for scientific knowledge production}
Although outsiders are often more open to adaptation and capable of bringing fresh perspectives to new fields, they also face distinct challenges, particularly in recognition and the lack of established expertise. \citet{Hill2021-ev} introduce the concept ``pivot penalty,'' highlighting the challenges researchers face when moving away from their established expertise. It finds that huge changes can lead to decreases impact, suggesting that outsiders can bring fresh perspectives but face difficulty in receiving recognition. Similar dynamics are observed in emerging interdisciplinary study around AI. For example, \citet{Tao2021-zf} analyzed the demographics of researchers at the intersection of machine learning and societal issues and found that individuals from underrepresented backgrounds (insiders) are more likely to pursue work on fairness, accountability, and transparency. 

\vspace{0.2cm}
\noindent In sum, outsiders can drive innovation by challenging dominant paradigms, breaking from conventions, and drawing on diverse social capital. But their position outside traditional structures also exposes them to constraints, such as lack of recognition, fewer resources, or gaps in foundational knowledge, that can limit their long-term influence. Drawing on the strengths of the insider–outsider perspective, which highlights the complexities of scientific collaboration, we adopt a framework originally proposed by \citet{Gibbons1994-kp} and later adapted by \citet{Hessels2008-ew} to analyze how AI shapes participation, knowledge negotiation, and innovation within and across research communities. We introduce our methodological approach in the following section.




\section{Method} \label{sec:Method}

\subsection{Adapting Insider-Outsider Perspective on Evaluating LLM's Scientific Impact}

To examine how AI is reshaping scientific knowledge production, we build on prior large-scale evaluations (macro-level evaluations described in Section \ref{subsec:Evaluation}) by incorporating an insider–outsider framework that highlights the socially situated and collaborative focus. Focusing on LLM as a case study, we examine how researchers’ disciplinary backgrounds and institutional affiliations shape their scientific knowledge production around LLMs. 

Based on the insider-outsider perspective described in Section \ref{subsec:Insider-outsider}, we distinguish between "insiders", such as computer scientists and AI researchers, who are directly involved in building, fine-tuning, and advancing LLM technologies, and "outsiders",
who apply LLMs within specific domain contexts, such as medicine, biology, or psychology, often without direct involvement in the technical development. This perspective allows us to move beyond aggregate trends towards investigating how the diffusion of LLM interacts with researchers' social and epistemic positions (e.g., transitions from insider to outsider or vice versa) and collaborative dynamics (e.g., collaborations between insiders and outsiders). 

We primarily draw on the theoretical framework of Mode 1 and Mode 2 knowledge production proposed by \citet{Gibbons1994-kp} and adapted by \citet{Hessels2008-ew}. This framework outlines key differences between two modes of producing knowledge across five dimensions in application focus, disciplinary orientation, collaboration patterns, accountability, and evaluation criteria. A detailed comparison of the two modes across these dimensions is presented in Table \ref{tab:knowledge comparision}. 

In the following sections, we first introduce dataset used to perform our large empirical evaluation (Section \ref{subsec:Dataset}). We then provide a detailed workflow (Section \ref{subsec:Workflow}), describing how we adapt this framework into computational methods to quantify AI's scientific impacts.

\begin{table}[ht]
\centering
\caption{Comparison Between Mode 1 and Mode 2 Knowledge Production}
\label{Table1}
\begin{tabular}{@{}>{\raggedright\arraybackslash}p{2cm}>{\raggedright\arraybackslash}p{2cm}p{8cm}@{}}
\toprule
\textbf{Mode 1} & \textbf{Mode 2} & \textbf{Differences} \\
\midrule
Academic context & Application context & 
Mode 1: Knowledge is generated for academic purposes, with a gap in time and space between discovery and application. \newline
Mode 2: Knowledge is produced in direct response to real-world problems, often with immediate use in mind. \\
\addlinespace
Disciplinary & Trans-disciplinary & 
Mode 1: Research is done within a single discipline. \newline
Mode 2: Research integrates theories and methods from multiple disciplines. \\
\addlinespace
Homogeneity & Heterogeneity & 
Mode 1: Research is conducted in homogeneous academic settings. \newline
Mode 2: Knowledge is generated in diverse environments (e.g., universities, industry, government, NGOs), reflecting broader participation. \\
\addlinespace
Autonomy & Reflexivity / Social Accountability & 
Mode 1: Researchers are autonomous and accountable mainly to the scientific community. \newline
Mode 2: Researchers are aware of and responsible for societal consequences, often involving stakeholder collaboration. \\
\addlinespace
Traditional Quality Control & Novel Quality Control & 
Mode 1: Research is evaluated by peer review from disciplinary experts. \newline
Mode 2: Quality is judged by a broader group (e.g., users, communities), emphasizing impact, usability, and social relevance alongside scientific rigor. \\
\bottomrule
\end{tabular}
\label{tab:knowledge comparision}
\end{table}

\subsection{Dataset} \label{subsec:Dataset}

\subsubsection{Dataset description and search scope} 
OpenAlex is a comprehensive scientific knowledge graph that includes data on papers, authors, publication venues, institutions, and disciplinary information  \citep{Priem2022-xb}. The disciplinary structure in OpenAlex is hierarchical, spanning six levels, with 19 root disciplines such as computer science, medicine, and biology. Because OpenAlex is updated monthly, it is suitable for tracking the latest developments in fast-moving research areas. 

To study emerging work on LLMs and ChatGPT, we use OpenAlex to identify preprints and articles published between 2023 and 2025 that contain the keywords ``large language model'' or ``chatgpt'' in their titles or abstracts. This yields a dataset of 32,167 papers until 2025 April. 

\subsubsection{Paper inclusion and exclusion criteria}
Although OpenAlex provides author disambiguation, its quality remains limited. To address this, we treat authors affiliated with the same institution as a single unique researcher. While this approach may result in some data loss, for example, when researchers move between institutions, it does not compromise the focus of this study. Our goal is to compare each researcher's LLM-related paper with their recent most similar-topic work to observe changes over time. 

Given the importance of authors' institution here, we removed those researchers without identified institution information. We also excluded LLM papers with titles containing terms like Review, Survey, or Summary, as these works primarily conduct literature overviews rather than adopting LLMs in their research. Such papers fall outside the scope of our study, since we aim to examine how LLMs reshape researchers’ prior research focus. 

Finally, we identified 7,106 unique researchers along with their LLM-related papers and corresponding prior papers on the most similar topics. There are 2,998 insiders and 4,108 outsiders. For outsiders, their primary fields of publication include medicine (1,588), psychology (494), biology (351), political science (80) and chemistry (74).

\subsection{Workflow} \label{subsec:Workflow}
\begin{figure}[ht]
    \centering
    \includegraphics[width=\textwidth]{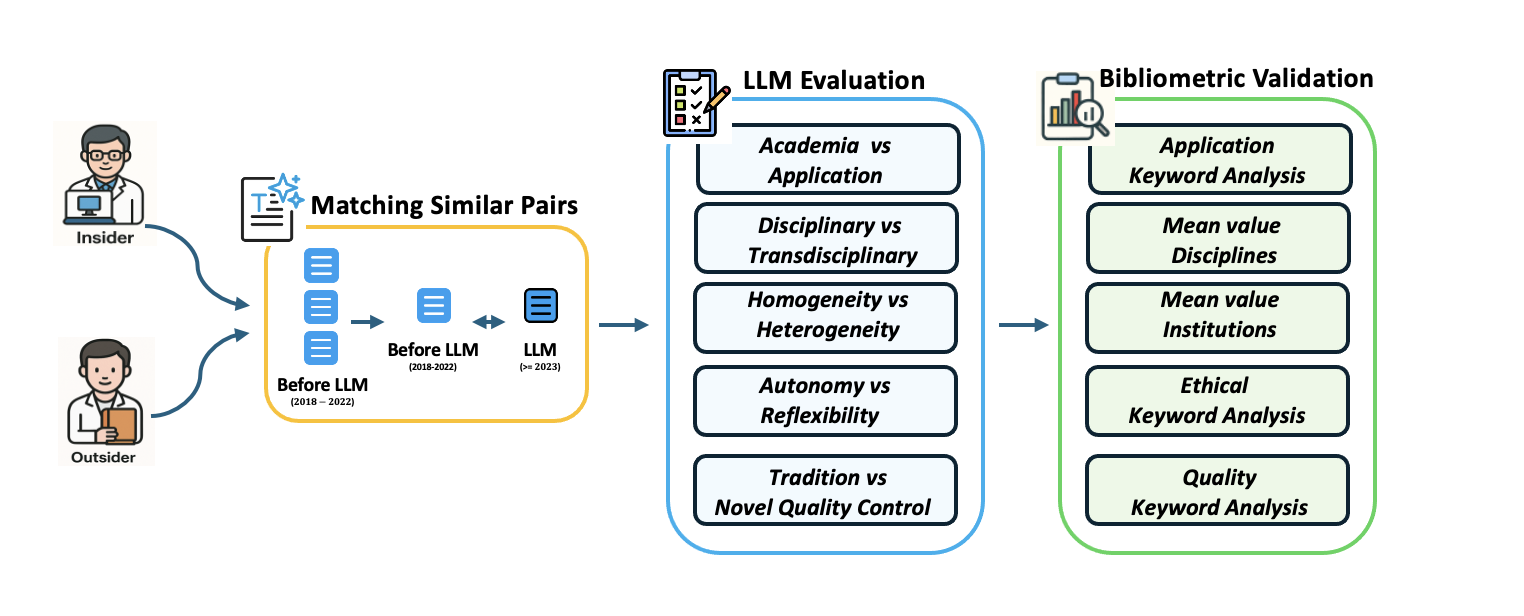}
    \caption{The computational process of evaluating the impact of LLM in scientific knowledge production}
    \Description{The computational process of evaluating the impact of LLM in scientific knowledge production}
    \label{fig:workflow}
\end{figure}

Figure \ref{fig:workflow} outlines the major computational process to evaluate the impact of LLM in scientific knowledge production. We now introduce each step in this section.

\subsubsection{Defining insiders and outsiders} 
\begin{figure}[ht]
    \centering
    \includegraphics[width=\textwidth]{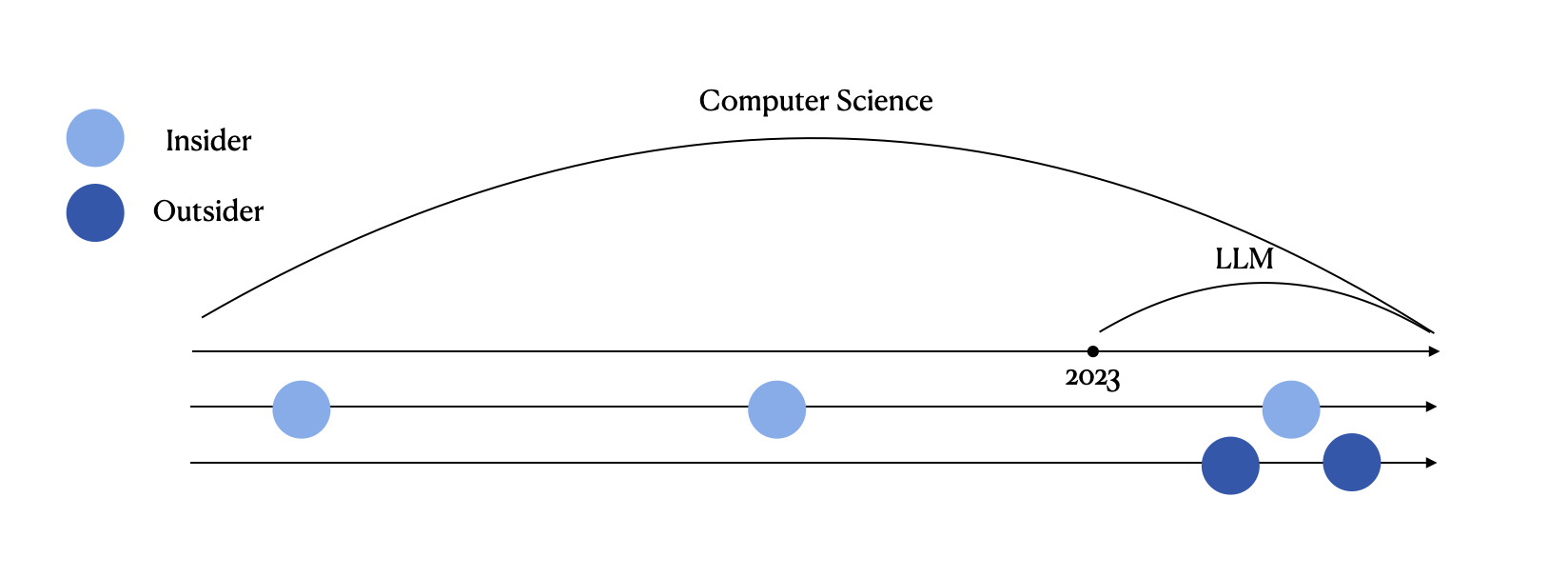}
    \caption{The definition of insider and outsider}
    \Description{The definition of insider and outsider}
    \label{fig:insider-outsider}
\end{figure}

We define insiders as researchers whose primary area of publication is in the field of \textit{Computer Science}. To determine this, we analyze each researcher's root-level disciplines from their publications between 2018 and 2022, the five years preceding the release of ChatGPT in 2023. The discipline with the highest frequency among their publications is considered their main research field. If this primary discipline is computer science, the researcher is classified as an insider (shown in light blue in Fig \ref{fig:insider-outsider}); otherwise, they are classified as an outsider (shown in dark blue).

Given the broad scope of computer science, we refined our classification using second-level disciplinary fields to more accurately identify "insiders." Instead of labeling all computer scientists as insiders, we focus on those working in subfields where LLM development is central, specifically, artificial intelligence (AI) and natural language processing (NLP). These subfields account for 91\% of the authors classified as insiders in our dataset, indicating that our operational definition aligns well with communities actively involved in the creation or fine-tuning of LLMs. 

In contrast, we define "outsiders" as researchers who apply LLMs in their work but are not directly engaged in building or fine-tuning these models. This group includes: (1) outsiders within computer science (e.g., systems, theory, or semantic web researchers); (2) outsiders from the natural sciences (e.g., medicine, biology); and (3) outsiders from the social sciences and humanities. This layered definition ensures a more precise and meaningful distinction between those contributing to the technical development of LLMs and those adopting them for domain-specific applications.

\subsubsection{Matching similar pairs between LLM papers and pre-LLM papers}
To assess the impact of LLMs on researchers’ knowledge production, we match each LLM-related paper with the most similar-topic paper authored by the same researcher in the five years prior to 2023. This allows us to compare the difference of knowledge production on similar topics before and after the rise of LLMs. To measure topic similarity, we use sentence embeddings generated by the Sentence-Transformers library \citep{Reimers2019-pc}. Specifically, we apply the pre-trained 'all-mpnet-base-v2' model to convert each abstract into a dense vector that captures its semantic meaning. We then compute cosine similarity between these vectors and select the highest-scoring pair for each researcher, linking their LLM paper with the most similar pre-LLM paper.

\subsubsection{Automating abstract classification using the knowledge framework with LLM evaluation} 

Based on the knowledge framework described in Table \ref{tab:knowledge comparision}, we use a language model prompted to act as an expert in the sociology of science and evaluate the difference mainly based on the paper abstract. Specifically, we construct a system prompt with several few-shot examples \citep{Brown2020-nj} as follows:

\begin{quote}
\begin{lstlisting}
You are an expert in the sociology of science. Your task is to classify research abstracts according to five dimensions that distinguish between Mode 1 and Mode 2 knowledge production. Each dimension is defined below with clear differences between Mode 1 and Mode 2. For each dimension, assign a label: 
1 = Mode 1, 
2 = Mode 2, 
0 = Not enough information to determine. 

Here are a few labeled examples to guide your classification:
\end{lstlisting}
\end{quote}
This prompt is used to evaluate each research abstract along the five dimensions. We implement this classification using the GPT-3.5-Turbo API, setting the temperature to 0 to ensure consistent, and context-aware responses. 

To guide the model’s classification, we selected two pairs of papers from our dataset as the few-shot examples. Each pair was authored by the same researcher on the same research topic but one is a pre-LLM paper and another is an LLM-related paper. To improve the few-shot accuracy, we annotate and provide human rationales to contextualize the knowledge shift from mode 1 to mode 2 (as described in Table \ref{tab:knowledge comparision}). We report our annotated rationales as follows.

Two research papers from the linguistics domain examine gender bias from different angles: one investigates gender representation in linguistic example sentences, while the other analyzes how LLMs exhibit gender-based occupational stereotypes:

\noindent
\begin{minipage}[t]{0.48\textwidth}
\vspace{0pt}
\begin{tcolorbox}[colback=gray!10, colframe=gray!50, boxrule=0.5pt, sharp corners]
\small
\textit{Gender biases in linguistic examples} \citep{Kotek2020-gw}\\ [2pt]
... We examine the articles published over the past 20 years in Language, Linguistic Inquiry, and Natural Language Linguistic Theory, and find striking similarities to this prior work ... We show that female-gendered arguments are less likely to be referred to using pronouns and are more likely to be referred to using a kinship term, whereas male-gendered arguments are more likely to have occupations and to perpetrate violence ...
\end{tcolorbox}
\end{minipage}
\hfill
\begin{minipage}[t]{0.48\textwidth}
\vspace{0pt}
\begin{tcolorbox}[colback=gray!10, colframe=gray!50, boxrule=0.5pt, sharp corners]
\small
\textit{Gender biases in LLMs} \citep{Kotek2023-nq}\\ [2pt]
... \textcolor{orange}{This paper investigates LLMs' behavior with respect to gender stereotypes} ... \textcolor{red}{We test four recently published LLMs and demonstrate that they express biased assumptions about men and women's occupations.} Our contributions in this paper are as follows: (a) ... (b) \textcolor{olive}{these choices align with people's perceptions better than with the ground truth as reflected in official job statistics} ... \textcolor{violet}{As with other types of societal biases, we suggest that LLMs must be carefully tested to ensure that they treat minoritized individuals and communities equitably.}
\end{tcolorbox}
\vspace{2pt}
\end{minipage}

We labeled the left example as Mode 1 knowledge production across all five dimensions, as it reflects research conducted within an academic context, rooted in a single discipline, situated in a homogeneous scholarly setting, accountable primarily to the scientific community, and evaluated through traditional quality control peer review. In contrast, the right example reflects four key dimensions for Mode 2 knowledge production: it is application-focused (highlighted in \textcolor{red}{red}), integrates methods from multiple disciplines (highlighted in \textcolor{orange}{orange}), demonstrate awareness of and responsibility for societal consequences (highlighted in \textcolor{violet}{violet}), and is reviewed by broader set of stakeholders beyond the academic community (highlighted in \textcolor{olive}{olive}). Since the abstract in this example does not specify whether the work extends beyond an academic setting, it is classified as exhibiting homogeneity rather than heterogeneity.

\subsubsection{Validating with bibliometrics measurements}
Although we use GPT-3.5-Turbo to classify papers across five dimensions application focus, disciplinary orientation, collaboration patterns, accountability, and evaluation criteria, the accuracy of the results is not guaranteed. Therefore, we conducted a corresponding bibliometric analysis to validate each dimension as described in Table \ref{tab:knowledge comparision}:

\begin{description}
    \item [Application Focus:] To assess the extent of application-oriented work, we searched for keywords related to applied contexts: ‘application’, ‘applied’, ‘practical’, ‘implementation’, ‘deployment’, ‘industrial’, ‘clinical’, ‘prototype’, ‘solution’, ‘adoption’, ‘utility’. The presence of these terms was used to distinguish academic context from application context.
    \item [Disciplinary Orientation:] We calculated the mean number of disciplines per paper to evaluate the degree of transdisciplinarity. 
    \item [Collaboration Patterns:] We measured the number of unique institution types associated with each paper as a proxy for the heterogeneity of collaboration. Institution types were drawn from the OpenAlex dataset and include university, company, archive, facility, healthcare, government, and others.
    \item [Accountability Focus:] To identify socially responsible research, we searched for keywords indicating ethical and accountability concerns: ‘ethics’, ‘ethical’, ‘bias’, ‘fairness’, ‘transparency’, ‘accountability’, ‘responsibility’, ‘trust’, ‘justice’, ‘privacy’.  These terms help distinguish between academic autonomy and social accountability.
    \item [Evaluation Criteria:] We analyzed evaluation-related terms to distinguish traditional from emerging quality control approaches: ‘evaluation’, ‘assessment’, ‘validation’, ‘robustness’, ‘reliability’, ‘accuracy’, ‘consistency’, ‘benchmark’, ‘reproducibility’, ‘precision’.
\end{description}

\section{Results} \label{sec:Results}


We report our quantitative findings derived from the evaluation workflow described in the Method (Section \ref{subsec:Workflow}). These findings span five key dimensions outlined in Table \ref{tab:knowledge comparision}: including application focus (Section \ref{subsec:application}), disciplinary orientation (Section \ref{subsec:disciplinary}), collaboration patterns (Section \ref{subsec:collaboration}), accountability focus (Section \ref{subsec:accountability}), and evaluation criteria (Section \ref{subsec:evaluation}). For each dimension, we highlight publication differences between insiders and outsiders from our few-shot outcomes. 
Additionally, we complement the model results with bibliometric measurement validations.

\subsection{Outsiders Drive Application-oriented LLM Research} \label{subsec:application}
\begin{figure}[ht]
    \centering
    \begin{subfigure}[b]{0.48\textwidth}
        \centering
        \includegraphics[width=0.8\textwidth]{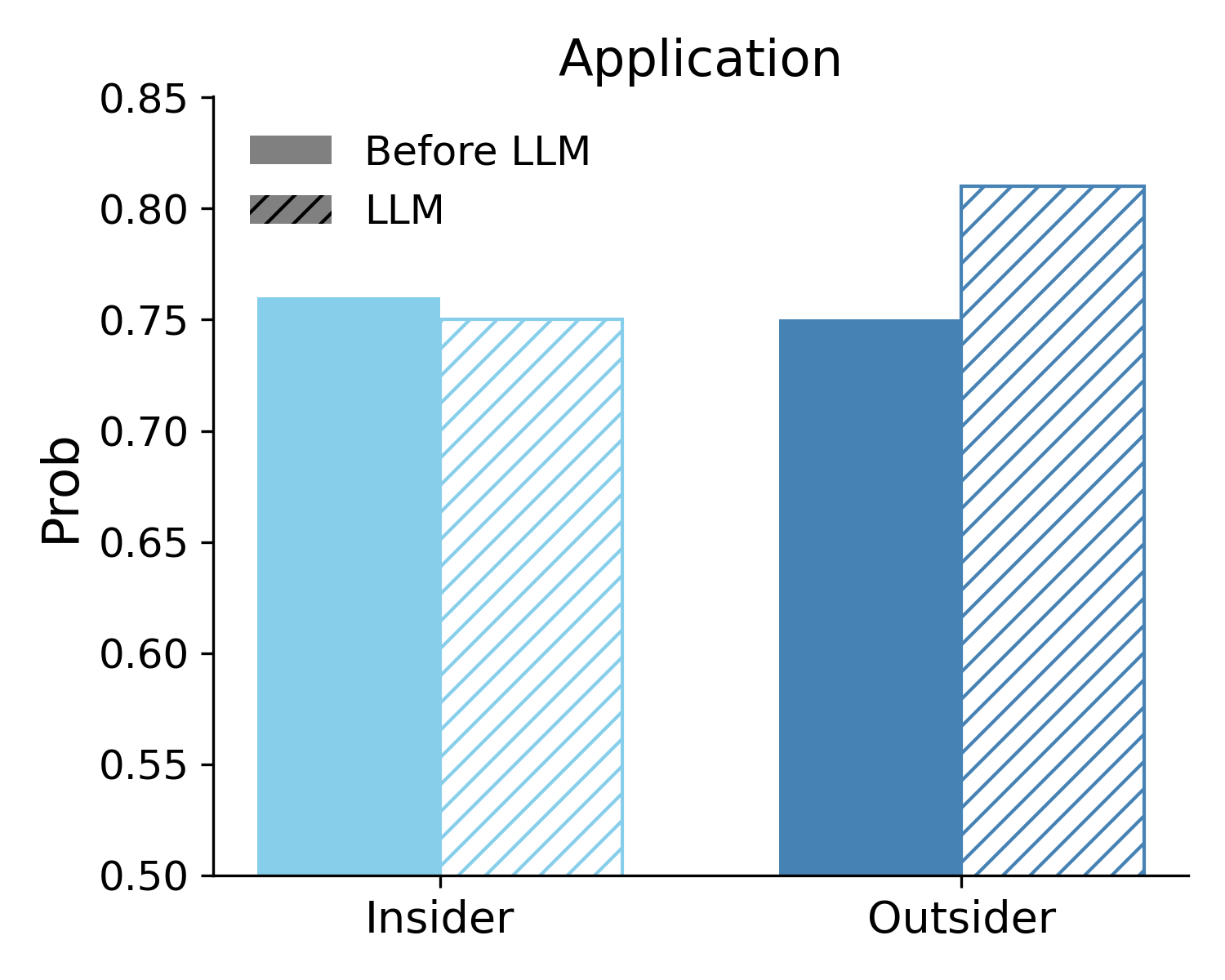}
        \caption{}
        \label{fig:app1}
    \end{subfigure}
    \hfill
    \begin{subfigure}[b]{0.48\textwidth}
        \centering
        \includegraphics[width=\textwidth]{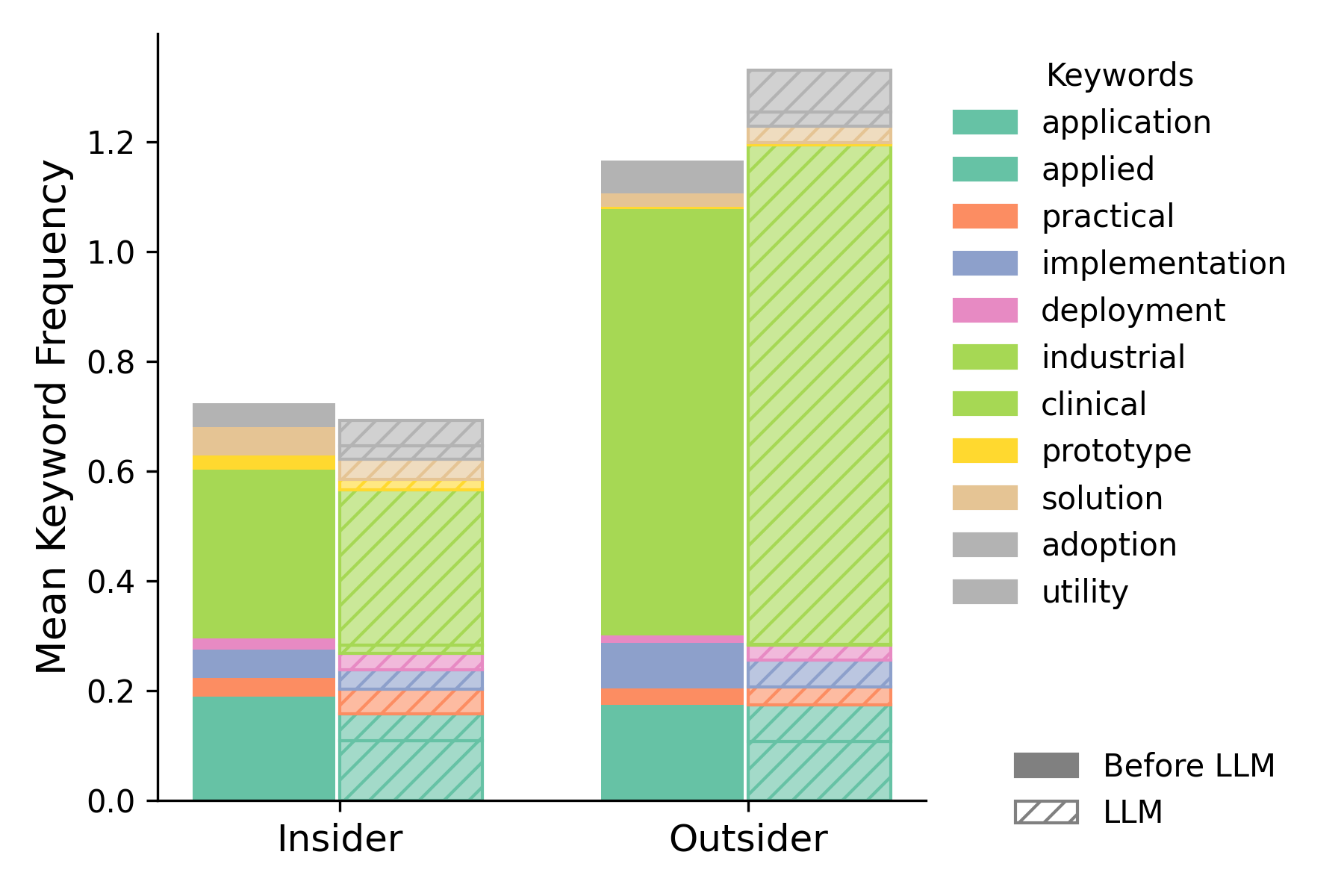}
        \caption{}
        \label{fig:app2}
    \end{subfigure}
    \caption{Comparison of application focus between pre-LLM and LLM papers for both insiders and outsiders. (a) Probability of being classified as application-oriented based on the model. (b) Distribution of application-related keywords across different groups.}
    \Description{Comparison of application focus between pre-LLM and LLM papers: (a) Model-based probability of application focus. (b) Keyword distribution across groups.}
    \label{fig:application focus}
\end{figure}

Based on the model's classification, we distinguish between academic and application-oriented contexts in LLM and pre-LLM papers. This allows us to compare how insiders and outsiders differ in their contextual focus after applying LLM in their papers. As shown in Fig \ref{fig:app1}, outsiders—researchers from non-computer science domains—have become notably more application-driven in their LLM-related work compared to insiders. Specifically, the proportion of application-oriented content among outsiders increased by 6\%, from 0.75 to 0.81. This suggests that outsiders tend to view LLMs as practical tools for solving domain-specific, real-world problems --- a trend further supported by our bibliographic analysis.

In our bibliographic analysis, we examined the frequency of application-related keywords. Consistent with the model’s findings, outsider papers show a greater increase in application-related terms, particularly keywords such as ``\textit{clinical}'' and ``\textit{utility}'', as shown in green and grey in Fig \ref{fig:app2}. This pattern aligns with prior findings suggesting that outsiders, such as those in psychology and health sciences, are increasingly leveraging LLMs to tackle domain-specific, applied challenges \citep{Ghafarollahi2024-in, Luo2025-ji, Sun2024-bn}. It also reflects a growing body of work in CSCW and HCI that focuses on developing LLM-enabled tools to support various research activities \citep{KapaniaShivani2025-wu, Liu2024-jo, Pang2025-eo}.
   
\subsection{Outsiders Expand Transdisciplinary Engagements with LLM} \label{subsec:disciplinary}

\begin{figure}[ht]
    \centering
    \begin{subfigure}[b]{0.48\textwidth}
        \centering
        \includegraphics[width=\textwidth]{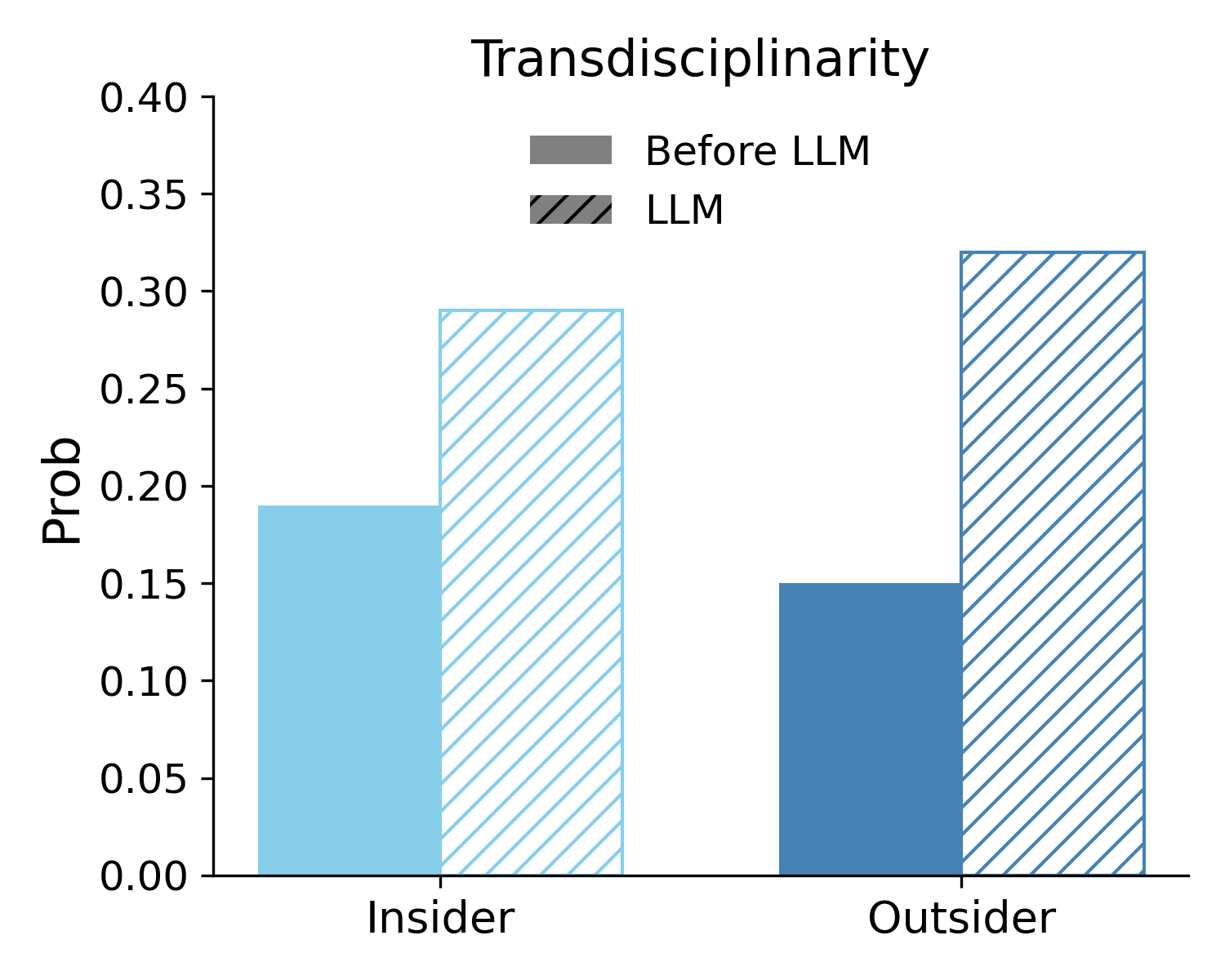}
        \caption{}
        \label{fig:dis1}
    \end{subfigure}
    \hfill
    \begin{subfigure}[b]{0.48\textwidth}
        \centering
        \includegraphics[width=\textwidth]{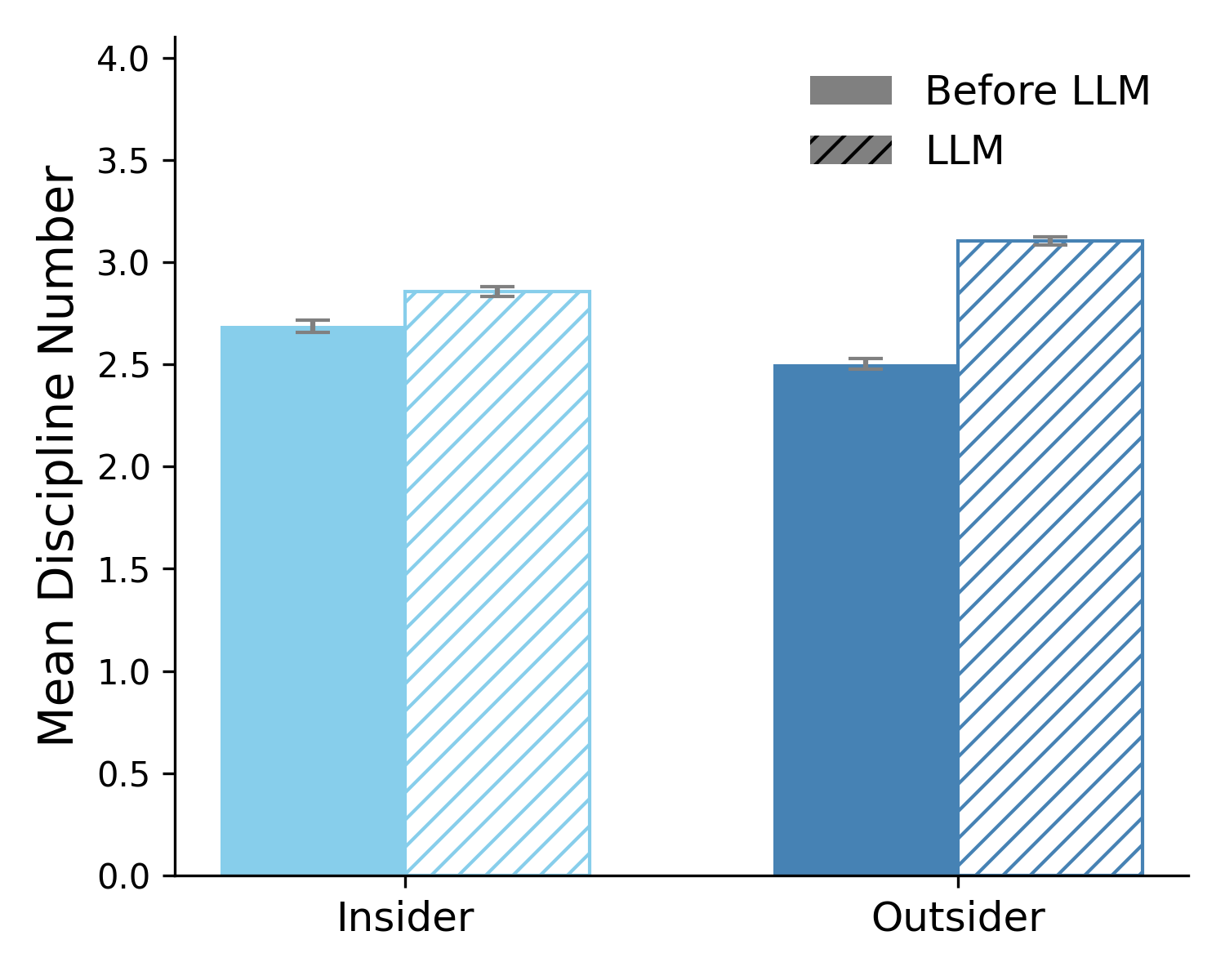}
        \caption{}
        \label{fig:dis2}
    \end{subfigure}
    \caption{Comparison of disciplinary orientation between pre-LLM and LLM papers: (a) Model-based probability of disciplinary orientation. (b) Keyword distribution across groups.}
    \Description{Comparison of disciplinary orientation between pre-LLM and LLM papers: (a) Model-based probability of disciplinary orientation. (b) Keyword distribution across groups.}
    \label{fig:disciplinary focus}
\end{figure}

Regarding disciplinary orientation—specifically whether LLM-related research is becoming more interdisciplinary—our few-shot classification experiments reveal a notable increase in transdisciplinarity among outsiders’ papers. The proportion of transdisciplinary work rose from 0.15 in pre-LLM papers to 0.32 in LLM papers (Fig \ref{fig:dis1} - deep blue). In contrast, insiders’ papers exhibit a more modest increase in transdisciplinarity, rising from 0.19 to 0.29 (Fig \ref{fig:dis1} - light blue). This suggests that, when incorporating LLMs, outsiders are more actively engaging in transdisciplinary research than their insider counterparts.

To further validate this finding, we calculated the average number of unique disciplines associated with each paper, based on their bibliographic metadata. Consistent with our model’s classification results, outsiders’ papers showed a notable increase in disciplinary diversity, rising from an average of 2.5 unique disciplines before the adoption of LLMs to 3.1 in the LLM era (Fig \ref{fig:dis2} - deep blue). In comparison, insider papers exhibited only a modest increase, from 2.69 to 2.86 (Fig \ref{fig:dis2} - light blue), indicating that outsiders are engaging with a broader range of disciplinary perspectives when incorporating LLMs into their research.
    
\subsection{Insiders Adapt LLM Research Collaborations Amid Growing Outsider Participation} \label{subsec:collaboration}
\begin{figure}[ht]
    \centering
    \begin{subfigure}[b]{0.48\textwidth}
        \centering
        \includegraphics[width=\textwidth]{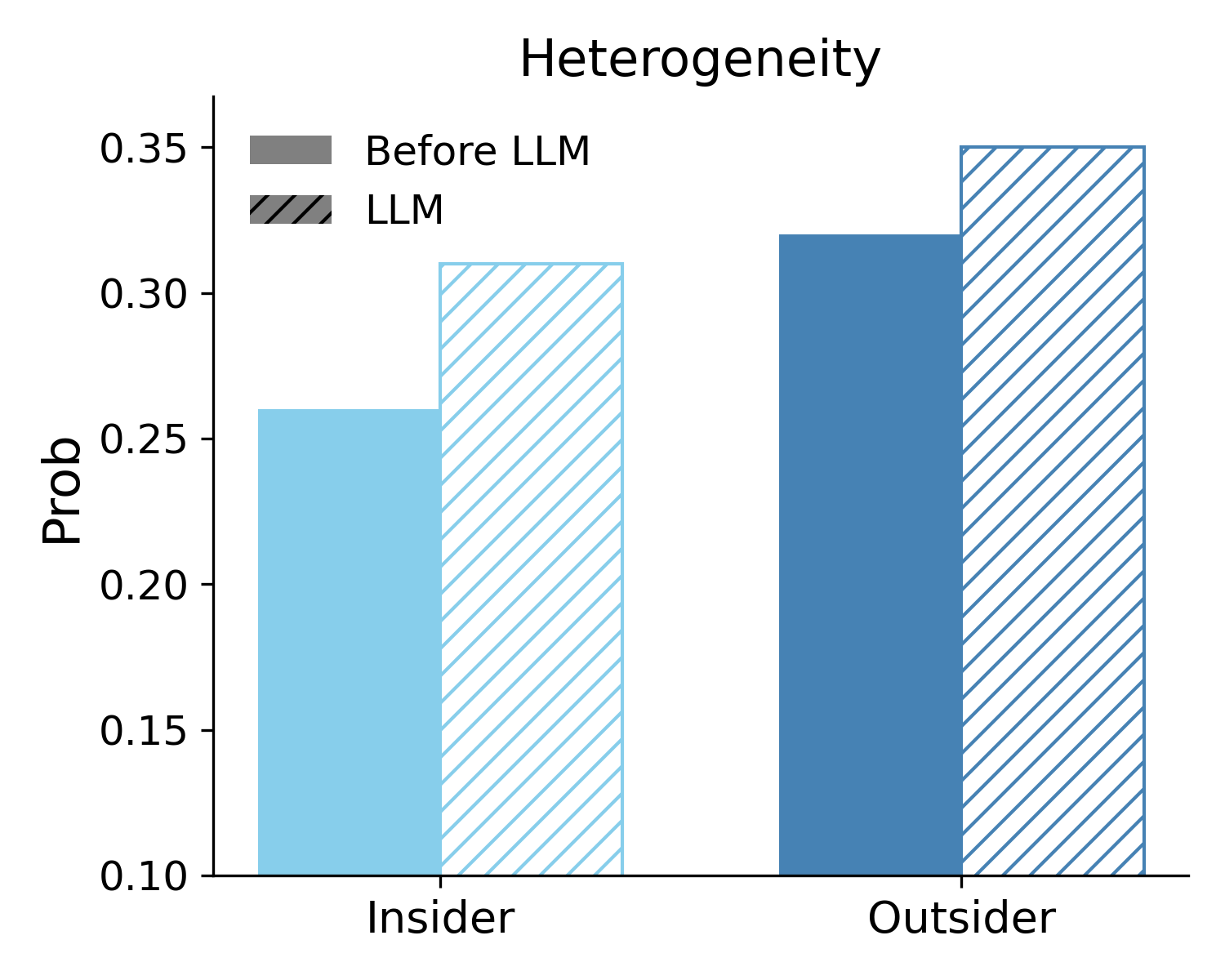}
        \caption{}
        \label{fig:hetero1}
    \end{subfigure}
    \hfill
    \begin{subfigure}[b]{0.48\textwidth}
        \centering
        \includegraphics[width=\textwidth]{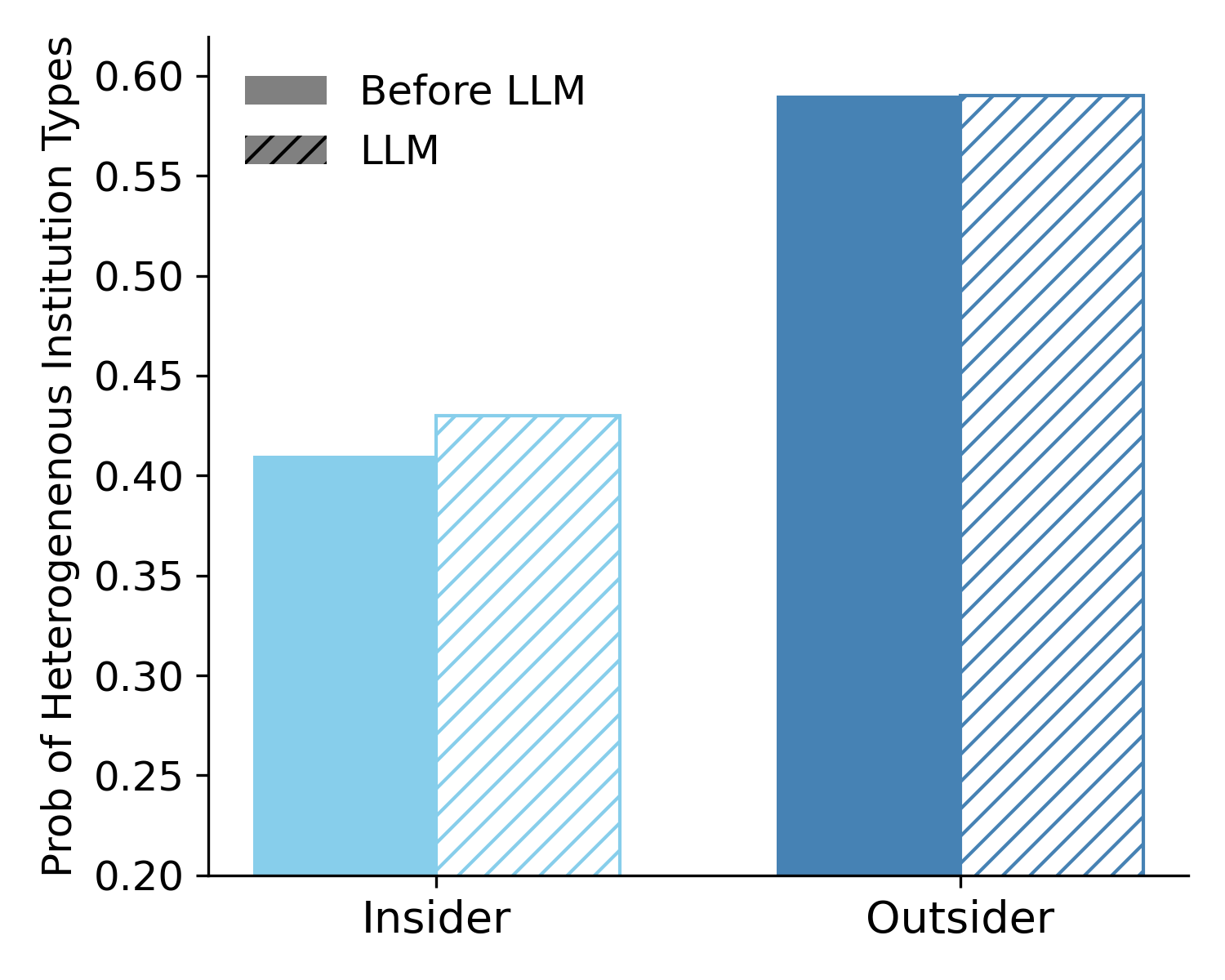}
        \caption{}
        \label{fig:hetero2}
    \end{subfigure}
    \caption{Comparison of collaboration patterns between pre-LLM and LLM papers for both insiders and outsiders. (a) Probability of being classified as having heterogeneous collaboration based on the model. (b) Probability of involving heterogeneous institution types (e.g., universities, industry, healthcare, government) across different groups.}
    \Description{Comparison of collaboration patterns between pre-LLM and LLM papers for both insiders and outsiders.}
    \label{fig:collaboration}
\end{figure}

We also examined the collaboration patterns of insiders and outsiders to assess whether knowledge production occurred within homogeneous academic settings or across diverse environments—an indicator of broader participation. Interestingly, and in contrast to prior findings, we observed a more pronounced increase in collaboration heterogeneity among insiders. As shown in Fig \ref{fig:collaboration}, the heterogeneity score for insider papers rose from 0.26 (pre-LLM) to 0.31 (LLM) (Fig \ref{fig:hetero1} - light blue). In comparison, outsider papers showed only a slight increase, from 0.32 to 0.35 (Fig \ref{fig:hetero2} - deep blue), suggesting that the shift toward more diverse collaborations was more significant among insiders during the LLM era.

To further validate this finding, we analyzed the probability of cross-institutional collaborations, specifically, whether co-authors were affiliated with different types of institutions (e.g., universities, healthcare organizations, companies, or government agencies). The results were consistent: in their LLM-era papers, insiders exhibited a higher likelihood of collaborating across institutional boundaries. 

However, it is important to note that despite this increase, insiders’ overall collaboration heterogeneity (0.31) remains lower than that of outsiders (0.35). One possible explanation for the relatively smaller change among outsiders is that LLM technologies have lowered the barrier to entry, enabling non-technical researchers to adopt LLMs within their existing collaboration networks without requiring significant restructuring. This interpretation might be partially supported by recent empirical studies, which show that researchers (especially for non-CS doctoral students) are increasingly using LLMs for programming tasks that previously demanded specialized technical expertise \citep{O-Brien2025-ex, Nam2024-sv}. As a result, their reliance on new technical collaborators could be potentially reduced, reinforcing the stability of their existing collaboration structures. In contrast, insiders may be actively restructuring their partnerships to preserve their domain distinctiveness and maintain a competitive edge, particularly as outsiders increasingly engage with and challenge traditional computer science domains.

\subsection{Outsiders Lead in Addressing Social Accountability of LLM Research} \label{subsec:accountability}

\begin{figure}[ht]
    \centering
    \begin{subfigure}[b]{0.48\textwidth}
        \centering
        \includegraphics[width=\textwidth]{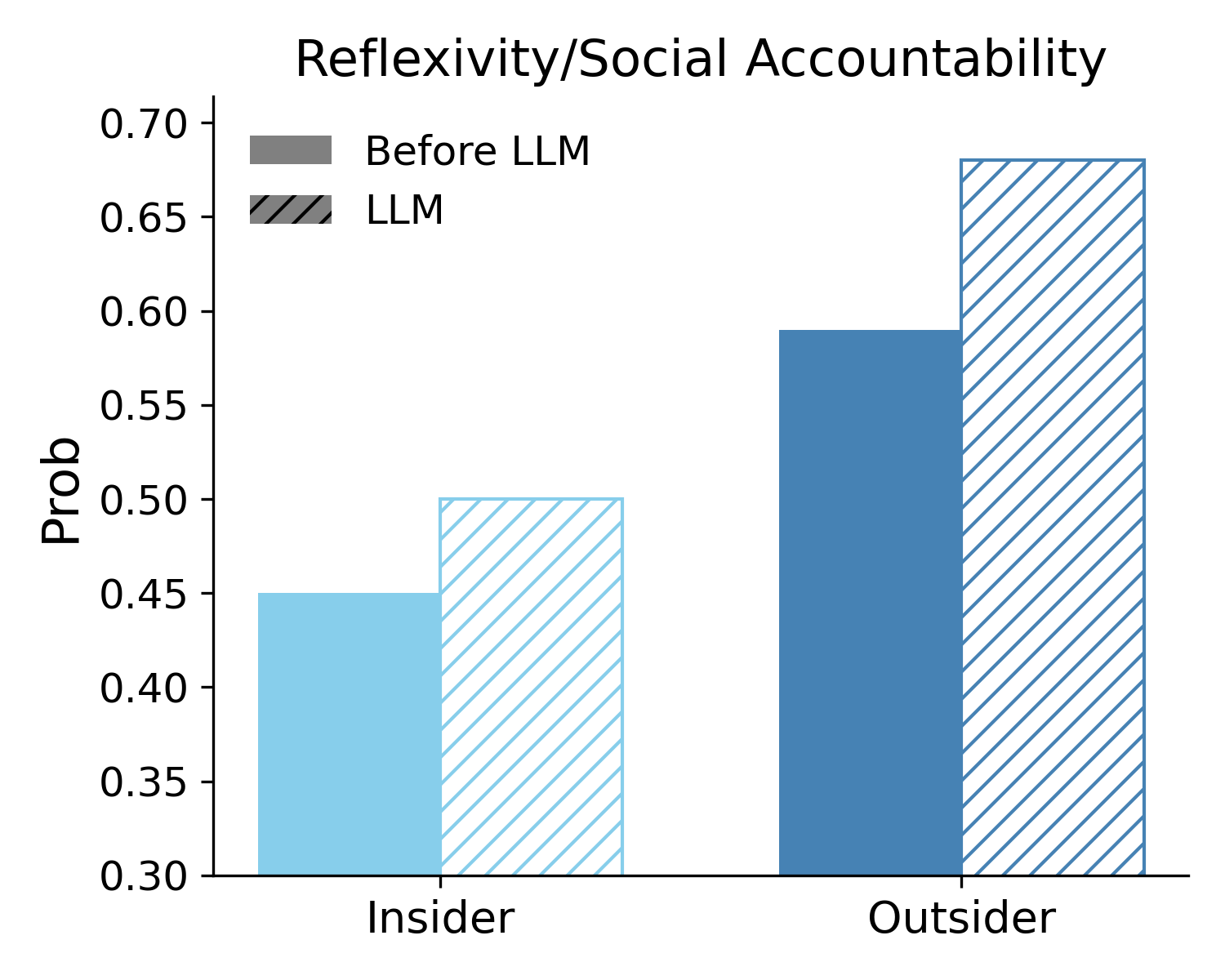}
        \caption{}
        \label{fig:acc1}
    \end{subfigure}
    \hfill
    \begin{subfigure}[b]{0.48\textwidth}
        \centering
        \includegraphics[width=\textwidth]{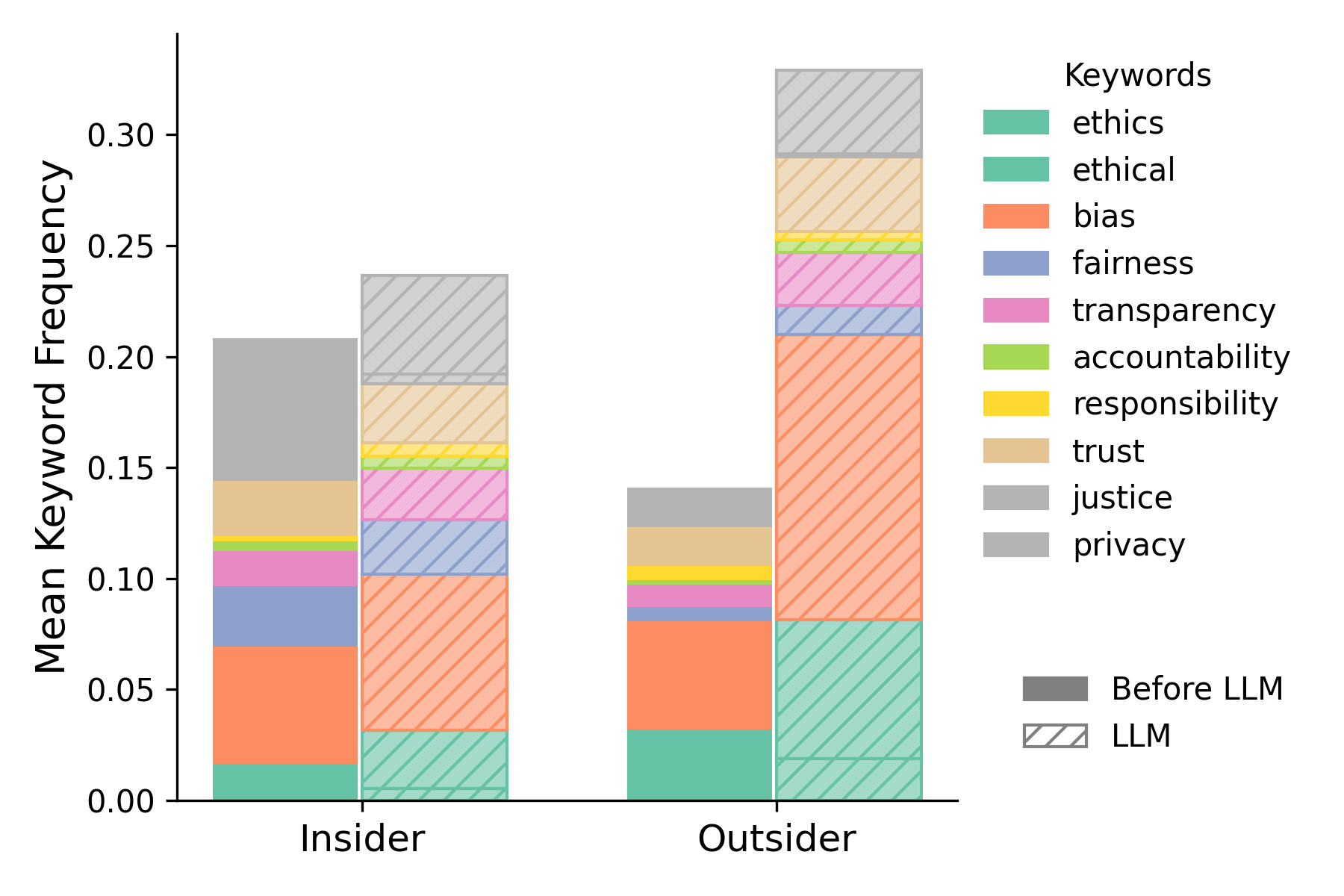}
        \caption{}
        \label{fig:acc2}
    \end{subfigure}
    \caption{Comparison of social accountability between pre-LLM and LLM papers for both insiders and outsiders. (a) Probability of being classified as social accountability based on the model (b) Distribution of accountability keywords across different groups.}
    \Description{Comparison of social accountability between pre-LLM and LLM papers for both insiders and outsiders.}
    \label{fig:accountability}
\end{figure}

In terms of the fourth dimension, social accountability, which concerns whether researchers operate autonomously within the scientific community or exhibit greater awareness of their broader social responsibilities, we observed a more substantial increase among outsiders. Specifically, the proportion of outsider papers reflecting social accountability rose from 0.59 to 0.68 after the adoption of LLMs (Fig \ref{fig:acc1}, deep blue). While insiders also showed an increase, it was more modest, rising from 0.45 to 0.50 (Fig \ref{fig:acc1}, light blue). These results suggest that, following the integration of LLMs into their work, outsiders are placing a greater emphasis on reflexivity and the societal implications of their research compared to insiders.

Supporting this, our bibliographic analysis (presented in Fig \ref{fig:acc2}) shows that although both insiders and outsiders exhibit an increase from pre-LLM to LLM papers, the rise is more pronounced among outsider disciplines. Outsider papers more frequently mention topics such as ethics, transparency, and bias, as reflected in the distribution of accountability-related keywords. These findings align with recent literature noting a growing emphasis on ethical awareness, fairness, and responsible AI practices in LLM-related research emerging from outsider disciplines \citep{Bail2024-pf, Stokel-Walker2023-dm, otis2024global}.

\subsection{Outsiders Broaden Evaluation Focus in LLM Research} \label{subsec:evaluation}

\begin{figure}[ht]
    \centering
    \begin{subfigure}[b]{0.48\textwidth}
        \centering
        \includegraphics[width=\textwidth]{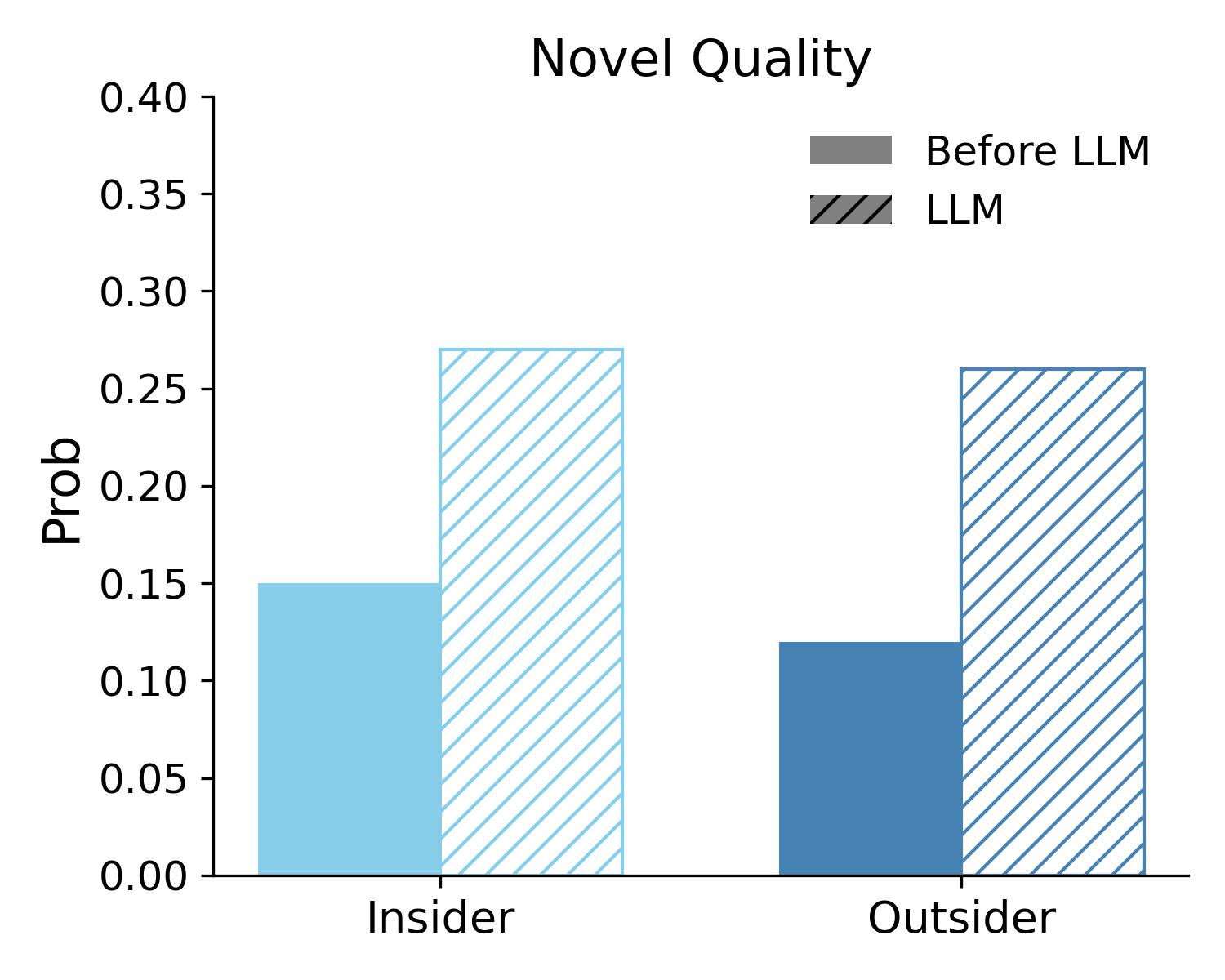}
        \caption{}
        \label{fig:eval1}
    \end{subfigure}
    \hfill
    \begin{subfigure}[b]{0.48\textwidth}
        \centering
        \includegraphics[width=\textwidth]{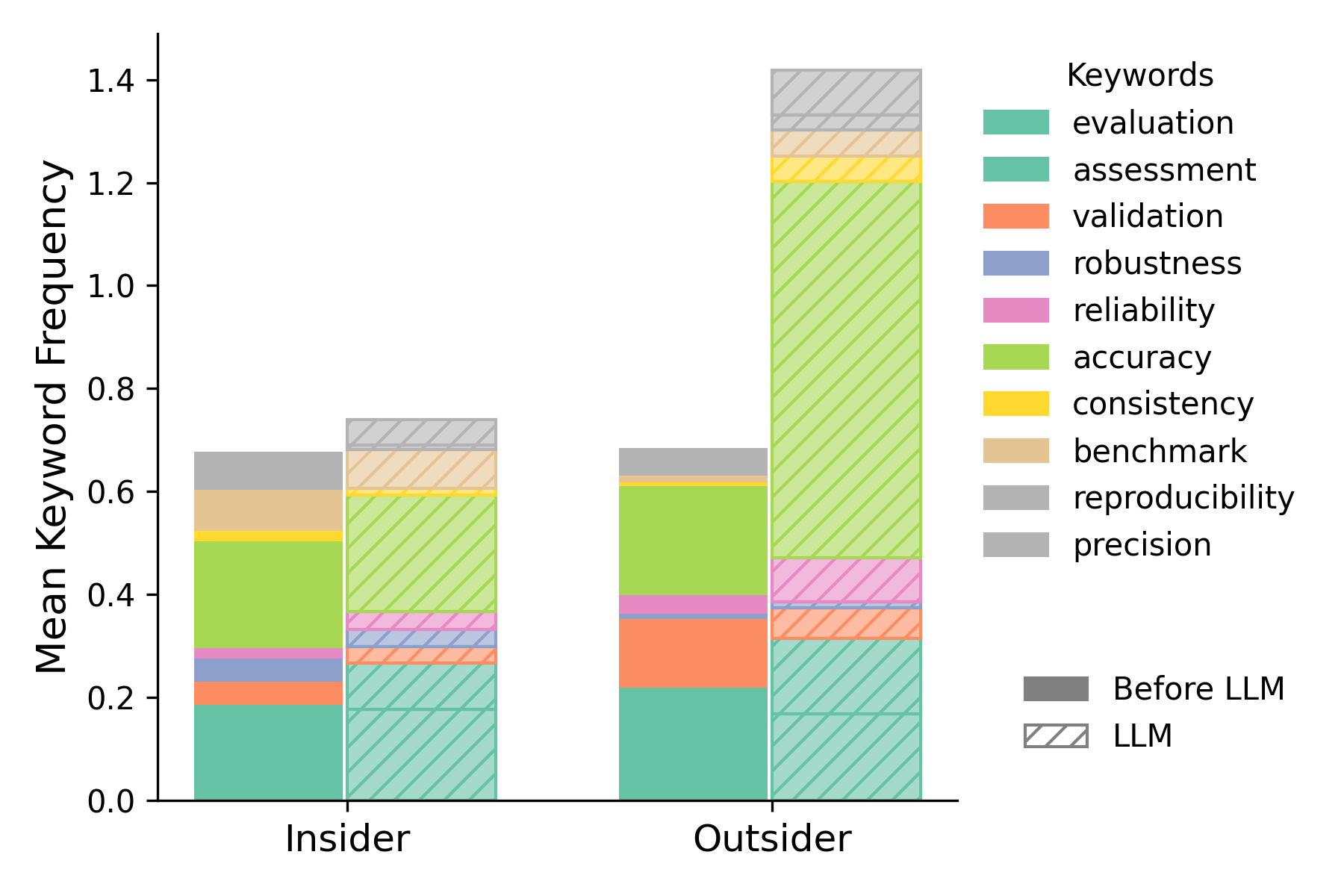}
        \caption{}
        \label{fig:eval2}
    \end{subfigure}
    \caption{Comparison of evaluation criteria between pre-LLM and LLM papers for both insiders and outsiders. (a) Probability of being classified as novel quality control based on the model (b) Distribution of evaluation keywords across different groups.}
    \Description{Comparison of evaluation criteria between pre-LLM and LLM papers for both insiders and outsiders.}
    \label{fig:evaluation}
\end{figure}

Regarding the final characteristic, evaluation criteria, which considers whether research is primarily assessed by disciplinary peers or by broader groups (e.g., users or communities) with an emphasis on social relevance in addition to scientific rigor, our few-shot results indicate that both insiders and outsiders exhibit notable shifts. However, the increase is more substantial among outsiders, suggesting a greater movement toward broader, socially oriented evaluation frameworks in their LLM-related research. Outsiders show an increase from 0.12 to 0.16 (Fig \ref{fig:eval1} - deep blue), while insiders display a rise from 0.15 to 0.27 (Fig \ref{fig:eval2} - light blue). 

However, our bibliographic analysis reveals a slightly different pattern. In the distribution of evaluation-related keywords, outsider papers show a more pronounced increase in mentions of terms such as accuracy, evaluation, precision, and reliability compared to their pre-LLM counterparts (Fig \ref{fig:eval2} - right). Notably, the magnitude of this increase is greater than what is reflected in the few-shot classification results. It is important to acknowledge a limitation here: these keywords may not exclusively refer to the evaluation of LLM effectiveness or novel quality control mechanisms. Instead, they could pertain to other aspects of the research. This ambiguity highlights an area that warrants further refinement in future analyses.

\section{Discussion} \label{sec:Discussion}

\subsection{Summary of Findings}


To examine how researchers' disciplinary and institutional positions shape their engagement with AI-driven research, we draw on a knowledge production framework (Table \ref{tab:knowledge comparision}) that distinguishes between insiders, those embedded within dominant AI fields, and outsiders, those from adjacent or non-CS domains. This framework is particularly instrumental for CSCW, which foregrounds the co-construction of sociotechnical systems through everyday practices, organizational roles, and institutional power structures that influence how knowledge is created, shared, and legitimized \citep{Ackerman2013-vc, Correia2019-yi, Jirotka2013-gt}.

We operationalized the knowledge framework through a combination of few-shot classification experiments and bibliographic analysis. The few-shot approach allowed us to probe shifts in research orientation across key epistemic dimensions (e.g., application-focus, transdisciplinarity, reflexivity), while the bibliographic analysis validated these findings by comparing AI-related publications authored by insiders and outsiders. Together, this mixed-method strategy reveals how researchers’ positions within or outside dominant structures mediate their capacity to shape emerging norms and practices in AI research.


By using LLM as an example, we found that as LLMs enter the research landscape, they appear to be reshaping the practices of outsiders more dramatically than those of insiders. In particular, researchers who were previously unaffiliated with the computer science field (i.e., outsiders) are adopting LLMs not only as technical tools but as catalysts for broader epistemic change. First, there is a marked shift toward application-oriented research, prioritizing knowledge that addresses real-world needs. Second, they expand the disciplinary boundaries of their work, engaging more frequently in transdisciplinary collaborations. Moreover, their LLM-related research increasingly reflects concerns around social accountability, including ethical transparency and reflexivity. Finally, outsiders also adopt more rigorous evaluation language, suggesting a sharpened focus on LLM reliability and evaluations.

In contrast, researchers already rooted in computer science (i.e., insiders) exhibit a subtler transformation. While they do not shift as significantly across most dimensions of knowledge production compared to outsiders, they show a notable increase in collaboration heterogeneity, partnering with a wider range of institutional actors. This pattern may signal a strategic repositioning, as insiders adapt to a research environment that is now shared with a growing number of LLM-enabled outsiders.


In sum, the above patterns suggest that outsiders are using LLMs as an entry point to reshape research practices, pushing toward more applied, transdisciplinary, accountable science with novel approaches to quality control. Meanwhile, insiders are responding with selective adaptation, particularly in seeking heterogenous collaborators, to preserve their positions within an evolving landscape.

\subsection{Expanding CSCW Research Focus to Support Outsider-led AI Research}

As our findings show, outsiders are increasingly adopting LLMs to address domain-specific challenges across fields such as health \citep{Li2024-vc, Naeem2024-ve}, psychology \citep{Almeida2024-ll}, social sciences \citep{Bail2024-pf}, and more. This trend presents a timely opportunity for the CSCW and HCI communities to reorient our research focus. Rather than concentrating primarily on LLM use within AI-related domains or our own disciplinary spaces, we argue for expanding both the scope of study and the design of tools to better support how LLMs are integrated into other research fields. This means moving beyond general-purpose research support toward the creation of systems that align with the specific tasks, workflows, and epistemic practices of diverse disciplines.

Currently, much of the work on LLM adoption in research contexts remains inward-facing. For example, recent studies have explored ethical concerns around LLM use in HCI research \citep{KapaniaShivani2025-wu}, the deployment of synthetic LLM personas \citep{Prpa2024-dx}, and the overall integration of LLMs into HCI research landscape \citep{Pang2025-eo}. Similarly, evaluations of new LLM-enabled tools designed for research activities tend to rely on feedback from participants within the HCI or AI communities \citep{Kang2023-yh, Pu2024-jr, Feng2024-xl}. While these studies are valuable, they reflect a limited view of how LLMs are reshaping knowledge production across different research domains.

We argue that expanding the scope of LLM research in CSCW and HCI to include other disciplines could offer two key benefits. First, researchers in domain-specific fields can provide critical insights into how LLMs succeed, or fail, to support the nuanced reasoning required in their areas of expertise. Although LLMs often perform well on general reasoning tasks, they may fall short in capturing the deep contextual knowledge, procedural logic, and validation standards central to many disciplines. Engaging with these researchers can help identify such gaps, informing not only future AI research but also the co-design of more robust and accountable systems. Notably, CSCW and HCI scholars are uniquely positioned to serve as intermediaries in advancing domain-specific AI applications, such as in journalism \citep{Petridis2023-dd, Nishal2024-oa, Liu2024-pr}, social media research \citep{Jia2024-zi}, and legal studies \citep{Cheong2024-cu, Delgado2022-ri}, where these fields are increasingly intersecting with AI technologies. Such interdisciplinary mediation is also likely to succeed in the context of scientific collaboration, where shared goals and complementary expertise can drive innovation.

Second, this shift opens up opportunities for designing novel LLM-enabled tools that are tailored to the unique workflows and knowledge practices of specific fields. For example, in astrophysics, researchers have developed agent-based LLM systems that support tasks such as galaxy image retrieval \citep{Mishra-Sharma2024-lh} and spectral model interpretation \citep{Sun2024-bn}, capabilities deeply grounded in the epistemic practices of the field. These types of domain-specific innovations underscore the creative design potential that arises when tool development is informed by situated expertise.

\subsection{Articulating the Moderated Work in Transdisciplinary Scientific Collaboration}


Our findings reveal a significant rise in transdisciplinary collaboration fueled by the adoption of LLMs. This shift is particularly evident in two forms: 1) outsider researchers increasingly working across disciplinary boundaries (Section \ref{subsec:disciplinary}) and 2) insiders actively restructuring their collaboration networks beyond academic settings (Section \ref{subsec:collaboration}). While some of these collaborations may emerge organically, many require active mediation. Note that differences in domain language, norms, and evaluation standards could make coordination challenging across diverse disciplines.

CSCW research has a long tradition of studying the mediating roles within sociotechnical systems, such as content moderators who navigate between users and platforms \citep{Steiger2021-xl, Li2022-fe}, or community managers who facilitate collective participation \citep{Kittur2008-ta, Kittur2010-yt}. As the AI-for-Science community continues to grow \citep{Van-Noorden2023-wp, Binz2025-bz}, we argue that analogous roles will become increasingly important and merit greater attention from CSCW scholars. 

Prior research on successful cross-domain collaboration highlights the importance of infrastructural and social mechanisms for bridging disciplinary divides \citep{Lyu2025-av, Young2017-vd}. Drawing on this, future CSCW research can contribute by investigating the \textit{Who}, \textit{What}, and \textit{How} of transdisciplinary collaboration in AI-driven science. This includes investigating: Who takes on the roles of fostering collaboration and managing data work (e.g., collecting domain-relevant dataset or curate contextual knowledge and use cases for AI applications); How participatory knowledge sharing practices are facilitated and sustained; and What kinds of technical infrastructures, potentially powered by AI, can be developed to support and scale such collaborative efforts across institutional and disciplinary boundaries.


\subsection{Study Limitations}
There are several limitations in this study. While we use GPT-3.5 classification to analyze knowledge production shifts, we currently do not prompt the model to explain its reasoning or provide evidence for its classifications. Although we attempt to validate the model's output using bibliometric measures, some inconsistencies remain. For instance, the keyword analysis may not exclusively reflect the impact of LLMs, instead these keywords could relate to other irrelevant aspects of research. This highlights the need for further refinement. In future work, we could prompt the model to justify its classifications by providing supporting evidence from the text, which could then be analyzed through keyword frequency. Additionally, incorporating human evaluation could help assess the relevance and accuracy of the model’s reasoning.

In addition, while our study focuses on large-scale patterns in LLM-related research with quantitative methods, it does not capture the nuanced mechanisms through which researchers adapt to these technologies. To complement our findings, we plan to conduct in-depth interviews with both insiders and outsiders in our dataset. These interviews will help illuminate the strategies, challenges, and motivations that shape how different communities integrate LLMs into their research practices.

Finally, our current definition of "outsiders" includes all researchers not primarily affiliated with core LLM-developing fields, but this category remains broad. This outsider group includes: (1) outsiders within computer science (e.g., systems, theory, or semantic web researchers); (2) outsiders from the natural sciences (e.g., medicine, biology); and (3) outsiders from the social sciences and humanities. It may differ significantly in how they engage with LLMs. We acknowledge this limitation and suggest that future work refine the classification of outsiders to better capture these important distinctions.

\section{Conclusion} \label{sec:Conclusion}
This study shows that the rise of generative AI, particularly LLMs, is reshaping scientific research practices. By analyzing changes in publication characteristics after LLM adoption, we find that both insiders and outsiders exhibit significant shifts across five key dimensions: research focus, disciplinary orientation, social accountability, collaboration, and evaluation criteria. These findings are based on automated LLM-driven classification and bibliometric validation. Outsiders, often from non-computer science fields, move toward more application-oriented, interdisciplinary, and socially engaged research, adopting new approaches to quality assessment. Meanwhile, insiders, researchers within computer science, respond by restructuring their collaborative networks. Together, these patterns suggest that LLMs are not only tools for innovation but also drivers of broader transformations in the way scientific knowledge is produced and organized.

\bibliographystyle{ACM-Reference-Format}
\bibliography{reference}


\pagebreak

\appendix

\end{document}